\documentclass[12pt,english,nofootinbib]{revtex4-2}
\usepackage{lmodern}

\usepackage[T1]{fontenc}
\usepackage[latin9]{inputenc}
\usepackage{babel}
\usepackage{array}
\usepackage{multirow}
\usepackage{dsfont}
\usepackage{amsmath}
\usepackage{amsthm}
\usepackage{amssymb}
\usepackage{graphicx}
\usepackage{geometry}
\usepackage[pdfusetitle,
bookmarks=true,bookmarksnumbered=false,bookmarksopen=false,
breaklinks=false, pdfborder={0 0 0},pdfborderstyle={},backref=false,
colorlinks=false]
 {hyperref}
 \usepackage[caption=false]{subfig}

\makeatletter

\usepackage{diagbox}
\usepackage{color}

\usepackage[normalem]{ulem}

\makeatother

\begin{document}
\title{Generating anisotropic models for relativistic stellar objects}
\author{Paulo Luz}
\email{paulo.luz@tecnico.ulisboa.pt}

\affiliation{Centro de Astrof\'isica e Gravita\c{c}\~ao - CENTRA, Departamento
de F\'isica, Instituto Superior T\'ecnico - IST, Universidade de
Lisboa - UL, Av. Rovisco Pais 1, 1049-001 Lisboa, Portugal}
\author{Sante Carloni}
\email{sante.carloni@unige.it}

\affiliation{Institute of Theoretical Physics, Faculty of Mathematics and Physics,
Charles University, Prague, V Hole\v{s}ovi\v{c}k\'ach 2, 180 00
Prague 8, Czech Republic}
\affiliation{DIME, Universit\`a di Genova, Via all'Opera Pia 15, 16145 Genova,
Italy,}
\affiliation{INFN Sezione di Genova, Via Dodecaneso 33, 16146 Genova, Italy}
\begin{abstract}
We introduce a new type of generating theorems in General Relativity for anisotropic, static, spherically symmetric solutions of the Einstein field equations. The results are used to derive a class of solutions that can serve as new models for the interiors of compact stars. Their geometric and thermodynamic properties are studied in detail, and we show that some of the new spacetimes contain, as particular cases, other well-known solutions. Focusing on a constant-density solution, we assess the relevance of the newly found geometry as a candidate for the incompressible limit of anisotropic compact stellar objects, comparing its features with those of the Bowers-Liang spacetime.
\end{abstract}
\maketitle

\section{Introduction}

It has been known for a long time that shear pressure plays a crucial role in the structure of compact stellar objects, most notably neutron stars, generating anisotropies able to influence their gravitational field \cite{Herrera:1997plx,Herrera_Barreto_2013,Herrera:2007kz,Chaisi:2006sc,Herrera:2015vca,Harko:2002pxr,Harko:2003pxr,Hoffberg:1970vqj,Rago:1991qe,Richardson:1972xn,Viaggiu:2008yf}. In spite of its role, the introduction of the additional degrees of freedom associated with shear stresses within matter has been comparatively less studied than the shearless case. 

The detection of gravitational waves~\cite{LIGOScientific:2016aoc}, and the dependency of their morphology on the details of the intimate structure of compact objects, has fostered a renewed attention on the properties of matter (see e.g. \cite{Cadogan:2024mcl,Cadogan:2024ohj,Cadogan:2024ywc}), and several new studies have been proposed, focusing on solutions of the Einstein field equations with anisotropic sources (see e.g. \cite{Thirukkanesh:2018hfy,Estrada:2018zbh,Singh:2016mqs,Maurya:2016oml,Maurya:2017gth,Ivanov:2017kyr,Jasim:2016cmk,Maurya:2016ecj,Bhar:2017ynp,Ovalle:2017fgl,Gabbanelli:2018bhs,Ovalle:2017wqi,Maurya:2015maa,Abellan:2020nkl}).

In general, the use of an appropriate formalism can substantially simplify the derivation of solutions. The 1+1+2 covariant formalism \cite{Clarkson_Barrett_2003,Clarkson_2007} has recently proven to be a powerful tool for the analysis, classification, and exploration of astrophysical objects \cite{Betschart_Clarkson_2004,Luz_Carloni_2024a,Luz_Carloni_2024b,Luz_Carloni_2024c}. In fact, within this formalism, one can derive a set of scalar equations that are equivalent to the Tolman-Oppenheimer-Volkov (TOV) equations \cite{Carloni:2017rpu,Carloni:2017bck}, but with a simpler and clearer mathematical structure. These equations also have the useful property of being easily generalized to include shear pressures.

A particularly interesting property of the TOV equations is contained in a set of theorems dubbed ``Generating Theorems''~\cite{Rahman,Lake, Martin,Boonserm:2005ni,Boonserm:2006up,ExactSolutions}. This set of results show that the solution-space of the TOV equations is structured into chains of solutions connected by specific transformations. The covariant formulation of the TOV equations enables us to recognize that generating theorems can be viewed as simple linear deformations of the variables that describe these solutions, and to discover, most importantly, that many types of deformations are possible, even those that mix the variables. This result, together with reconstruction approaches \cite{Carloni:2017rpu,Carloni:2017bck}, allows us to derive numerous exact anisotropic solutions of the TOV equations.

In this paper, we will focus on such type of theorems. We will use them to deduce properties of known static and spherically symmetric solutions and to derive a new class of solutions, further developing an idea initially proposed in \cite{Carloni:2017bck}.  We will then explore in detail the properties of the simplest element of this class of spacetimes, characterized by constant energy density, arguing that in some problems its use is preferable to the well-known Bowers-Liang solution.

The paper is structured as follows. In Section~\ref{Sec:1+1+2TOVeq}, the covariant formalism and the covariant TOV equations, as well as the generating theorem, are briefly presented. In Section~\ref{Unique_Sec}, we show how it is possible to use generating theorems to prove that the interior Schwarzschild solution is unique and that the original generating theorems cannot be used to obtain exact, analytical, physically relevant solutions for constant density stars.  In Section~\ref{NewSol_Sec}, we use a new type of generating theorem to derive a new class of solutions starting from the interior Schwarzschild metric. In Section~\ref{Sec:NewSol_Prop}, we investigate the properties of two simple members of this class of solutions, connecting the one characterized by a constant density source to other known anisotropic spacetimes. We also attempt a comparative assessment against the Bowers-Liang solution. Finally, in Section~\ref{Concl} we give our conclusions.

Throughout the article, we will work in the geometrized unit system
where $8\pi G=c=1$, and consider the metric signature $(-+++)$.

\section{1+1+2 Covariant TOV equations}\label{Sec:1+1+2TOVeq}
Let a static manifold with Locally Rotationally Symmetric (LRS) class II geometry, that is, a static spherically symmetric spacetime, permeated by a fluid that exhibits shearing stresses. An efficient way to treat this kind of geometry is to employ the so-called 1+1+2 covariant formalism~\cite{Clarkson_Barrett_2003,Clarkson_2007}. 

The 1+1+2 framework is constructed as follows. Let a timelike and a spacelike congruence characterized, respectively, by the tangent vector fields $u$ and $e$, pointwise orthogonal to each other and locally threading the spacetime. The 2-surfaces orthogonal to both $u$ and $e$ locally foliate the manifold. The geometry of the 2-surfaces is fully characterized by the tensor field
\begin{equation}
N_{ab}=g_{ab}+u_a u_b- e_a e_b=h_{ab}-e_a e_b\,,
\end{equation}
where $g_{ab}$ represents the components of the metric tensor in a local coordinate system, $u_a$  and $e_a$  the components of the 1-forms associated, respectively, to the vector fields $u$ and $e$. The tensor field $h$, with components $h_{ab}$, is called the projector onto the 3-surfaces orthogonal to $u$.

For a static LRS class II spacetime, we can define the differential operators
\begin{equation}
	\begin{split}
	&\dot{X}^{a...b}{}_{c...d}{} \equiv  u^{i} \nabla_{i} {X}^{a...b}{}_{c...d}\,, \\
	& D_{i}X^{a...b}{}_{c...d}{} \equiv  h^a{}_f ... h^b{}_g h^p{}_c ... h^q{}_d h^j{}_i \nabla_{j} {X}^{f...g}{}_{p...q}\,,\\
	& \widehat{X}^{a...b}{}_{c...d} \equiv  e^{i}D_i X^{a...b}{}_{c...d}~,  \\
	&\delta_i X^{a...b}{}_{c...d} \equiv  N^{a}{}_{f}...N^{b}{}_g N^{p}{}_{c}...
	N^{q}{}_{d} N^j{}_i D_jX^{f...g}{}_{p...q}\,, 
	\end{split}
\end{equation}
and considering the Weyl tensor, with components $C^{ab}{}_{cd}$,  we can  construct the scalar geometrical variables 
\begin{equation}
	\label{1p1p2_definitions_eq:Kinematical_variables}
	\begin{split}
	& \mathcal{A}=e_a \dot{u}^{a}\,,\\
	& \phi = \delta_a e^a\,, \\ 
	&\mathcal{E}=C^{ab}{}_{cd}u^{c}u^{d}\left[ e_ae_b - \frac{1}{2}N_{ab}\right]\,.
	\end{split}
\end{equation}
Additionally, given the stress energy tensor of the source fields, with components $T_{ab}$, we have the non-trivial projections
\begin{equation}
	\label{1p1p2_definitions_eq:Thermodynamic_variables}
	\begin{split}
	& \mu=T_{ab}{u}^{a}{u}^{b}\,,\\
	&p=\frac{1}{3}T_{ab}\left({e}^{a}{e}^{b}+N^{ab}\right)\,,\\
	& \Pi=\frac{1}{3}T_{ab}\left(2{e}^{a}{e}^{b}-N^{ab}\right)\,,
	\end{split}
\end{equation}
which fully characterize the thermodynamic properties of the matter source, such that
\begin{equation}
	T_{ab}=\left(\mu+p-\frac{1}{2}\Pi\right)u_{a}u_{b}+\frac{3}{2}\Pi\,e_{a}e_{b}+\left(p-\frac{1}{2}\Pi\right)g_{ab}\,.
\end{equation}
The scalar field $\mu$ represents the local energy density, $p$  the isotropic pressure, and $\Pi$  the anisotropic pressure, following from shearing stresses. In particular, the isotropic and anisotropic pressures are related to the radial and tangential components of the pressure by
\begin{equation}
	\begin{aligned}
		p_{r} & =p+\Pi\,,\qquad & p_{t} & =p-\frac{1}{2}\Pi\,.
		\label{1p1p2_definitions_eq:Radial_tangential_pressures_p_Pi_relations}
	\end{aligned}
\end{equation}
Using these quantities, we define the radial and the tangential speed of sound, respectively, as
\begin{equation}
c^{2}_{s,r}=\left(\frac{\partial p_{r}}{\partial\mu}\right)_{s},\quad c^{2}_{s,t}=\left(\frac{\partial p_{t}}{\partial\mu}\right)_{s},
\end{equation}
where the derivatives are taken at constant entropy.

A static LRS class II spacetime with vanishing cosmological constant can be fully described by the 1+1+2 potentials in Eqs.~\eqref{1p1p2_definitions_eq:Kinematical_variables} and \eqref{1p1p2_definitions_eq:Thermodynamic_variables}. Nonetheless, in the following, it will also be useful to write an expression for the Gaussian curvature, $K$, of the 2-surfaces in terms of those variables. Namely, we have
\begin{equation}
K= \frac{1}{3} \mu -\mathcal{E} - \frac{1}{2} \Pi + \frac{1}{4} \phi^{2}\,.
\end{equation}

The governing equations for all of the 1+1+2 variables read~\cite{Carloni:2017bck}
\begin{equation}\label{1112Eq}
	\begin{split}
	{\widehat p}+\widehat\Pi&= -\left(\frac{3}{2}\phi+\mathcal{A}\right)\Pi-\left(\mu+p\right)\mathcal{A}\,,\\
	\widehat\phi &= -\frac{1}{2}\phi^2 -\frac{2}{3}\mu-\frac{1}{2}\Pi-\mathcal{E}\,,\\
	\widehat{\mathcal{A}} &= -\left(\mathcal{A}+\phi\right)\mathcal{A} + \frac{1}{2}\left(\mu +3p\right)\,,\\
	\widehat{\mathcal{E}}-\frac{1}{3}\widehat\mu + \frac{1}{2}\widehat\Pi &=- \frac{3}{2}\phi\left(\mathcal{E}+\frac{1}{2}\Pi\right)\,,\\
	\widehat{K} &= -\phi K\,,
	\end{split}
\end{equation}
together with the constraint
\begin{align}
	\frac{1}{3}\left(\mu+3p\right)+\frac{1}{2}\Pi-\mathcal{A}\phi & =\mathcal{E}\,.
	\label{eq:Radial_Adiabatic_constraint_background}
\end{align}

Following Ref.~\cite{Carloni:2017bck},  we will consider a dimensionless parameter $\rho$,  such that the Gaussian curvature  is given by
\begin{equation}\label{GaussRho}
	K= \frac{e^{-\rho}}{
	K_0}\,,
\end{equation}
where $K_0$ is a constant. Then, defining the normalized variables
\begin{equation}
	\label{1p1p2_definitions_eq:Normalized_vars}
	\begin{split}
	&X=\frac{\phi_{,\rho}}{\phi}\,,
	\qquad Y= \frac{\mathcal{A}}{\phi}\,,
	\qquad \mathcal{K}=\frac{K}{\phi^2}\,,
	\qquad E=\frac{\mathcal{E}}{\phi^2}\,,\\
	& \mathbb{M}=\frac{\mu}{\phi^2}\,,
	\qquad \mathds{P}=\frac{\Pi}{\phi^2}\,,
	\qquad P=\frac{p}{\phi^2}\,,
	\end{split}
\end{equation}
equations~\eqref{1112Eq} can be used to formulate the classical TOV equations in covariant form as
\begin{equation}
	\label{1p1p2_definitions_eq:TOV_Gen}
	\begin{split}
		&P_{,\rho }+\mathds{P}_{,\rho }=P \left(\mathbb{M}-2
		   \mathds{P}-3 \mathcal{K}+\frac{7}{4}\right) +\mathds{P} \left(\mathbb{M}-3
		   \mathcal{K}+\frac{1}{4}\right)\\
		&~~~~~~~~~~~\,+\left(\frac{1}{4}-\mathcal{K}\right)
		   \mathbb{M}-P^2-\mathds{P}^2\,,\\
		&\mathcal{K}_{,\rho }=-2 \mathcal{K}\left(\mathcal{K}-\frac{1}{4}-\mathbb{M}\right)\,,\\
		&1- 4\mathcal{K}-4P+4Y-4\mathds{P}=0\,,\\
		&2 \mathbb{M}+2 P+2 \mathds{P}+2X-2 Y+1=0\,,\\
		&2 \mathbb{M}+6 P+3 \mathds{P}-6 Y-6 E=0\,,
	\end{split}
\end{equation}
where the comma represents derivative with respect to the indicated variable. 

The system~\eqref{1p1p2_definitions_eq:TOV_Gen} does not close. The properties of the matter fluid must be specified by an equation of state, and an equation that characterizes the source of the anisotropic stress. These relations usually follow from thermodynamic considerations of the source fluid, in turn derived from a fundamental matter description, e.g., quantum field theory or quantum chromodynamics, or an effective theory.

It can be shown that, by transforming the thermodynamic parameters of the source fluid, the solution space of the TOV equations has a structure. In particular, it is possible to map pairs of solutions, thus generating chains of solutions via specific transformations. This feature is the core of the so-called generating theorems and explains their name.

For example, in the isotropic case, $\Pi_\ast=0$, starting from the constant density, interior Schwarzschild solution, one can consider the transformation
\begin{equation}\label{AnzatzPY}
\begin{split}
P &=P_\ast+\bar{P}\,,\\
Y &=Y_\ast +\bar{Y}\,,
\end{split}
\end{equation}
and keep the remaining potentials fixed, namely $\mathcal{K}=\mathcal{K_\ast}$, $\mathbb{P}=\mathbb{P_\ast}=0$ and $\mathbb{M}=\mathbb{M_\ast}$. Here, and in what follows, an asterisk marks quantities that characterize an undeformed solution of the TOV equations and the barred variables represent a deformation to the original solution. Imposing that the new variables $P$ and $Y$ verify the TOV equations~\eqref{1p1p2_definitions_eq:TOV_Gen} and substituting in the deformation \eqref{AnzatzPY} yields
 \begin{equation}
\begin{split}
	\label{PY-theorem}
\bar{P}_{,\rho} &+\bar{P}^2+\bar{P} \left(3 \mathcal{K}_*-\mathbb{M}_*+2 P_*-\frac{7}{4}\right)=0\,,\\
\bar{Y}&=\bar{P}\,,
\end{split}
\end{equation}
The first equation is a Bernoulli equation and can be formally solved. The transformation~\eqref{AnzatzPY} and the resulting equations~\eqref{PY-theorem} can be called, for brevity, the ``$PY$ theorem''. Naturally, there is no guarantee that the new solution obtained by the application of this theorem will characterize a physically relevant setup.\footnote{The $PY$-theorem can also be extended to the case in which $\Pi_*\neq 0$ \cite{Carloni:2017bck}. In such an extension, the coefficient of the linear term in $\bar{P}$ of the first of Eqs. \eqref{PY-theorem} aquires a $2 \mathbb{P}_*$ term.}

Alternatively,  we can consider that the starting undeformed spacetime is permeated by a fluid with possibly non-trivial anisotropic stresses, and impose that the isotropic pressure of the deformed fluid is such that $P=P_\ast$. Under these hypotheses, a possible generating theorem, dubbed here ``$\mathbb{P}Y$ theorem'', follows from the deformation
\begin{equation}
	\label{AnzatzPPY}
\begin{split}
\mathds{P} &=\mathds{P}_\ast+\bar{\mathds{P}}\,,\\
Y &=Y_\ast +\bar{Y}\,,
\end{split}
\end{equation}
and keep the remaining potentials fixed. In this case, the TOV equations reduce to
 \begin{equation}
\begin{split}
	\label{PbbY-theorem}
\bar{\mathds{P}}_{,\rho} & +\bar{\mathds{P}}^2+\bar{\mathds{P}}
\left(3 \mathcal{K}_*-\mathbb{M}_*+2 \mathds{P}_*+2P_*-\frac{1}{4}\right)=0\,,\\
\bar{Y}&=\bar{\mathds{P}}\,,
\end{split}
\end{equation}

The $PY$ and the $\mathbb{P}Y$ theorems described above map solutions with a particular fluid source to solutions where one of the pressure terms is not changed. On the other hand, as detailed in Ref.~\cite{Carloni:2017bck}, another strategy to find solutions of the system is to construct a map from isotropic to anisotropic solutions.  Let us consider a concrete example, which will be called ``$P\mathds{P}$ theorem''. 

Let the pressure and the energy density of the new spacetime solution verify, $P=P_\ast-\mathds{P}$ and $\mu =\mu_\ast + \bar{\mu}$, where $\bar{\mu}$ is a generic function.\footnote{This transformation in reference \cite{Carloni:2017bck} contains a typographical error.}
That is, the normalized pressure variable of the isotropic solution, $P_\ast$, is equal to the normalized radial pressure of the new solution, $P+\mathds{P}$, and the difference between the energy density of the isotropic solution and the new solution is given by the function $\bar{\mu}$. In addition, let the isotropic and the new solution be characterized by the same Gaussian curvature, that is $K=K_\ast$, and  $\rho=\rho_\ast$.
These assumptions can be summarized by the transformation
\begin{equation}
	\begin{split}
		&P= P_\ast-\mathds{P}\,,\\
		&\frac{1}{\mathcal{K}}= \frac{1}{\mathcal{K}_\ast}+\frac{1}{\bar{\mathcal{K}}}
		\,,\\
		& \frac{\mathbb{M}}{\mathcal{K}}=\frac{\mathbb{M}_\ast}{\mathcal{K}_\ast} + \frac{\bar{\mathbb{M}}}{\mathcal{K}}\,.
		\label{1p1p2_definitions_eq:Ansatz_M_K}
	\end{split}
\end{equation}

Substituting Eq.~\eqref{1p1p2_definitions_eq:Ansatz_M_K} into  Eq.~\eqref{1p1p2_definitions_eq:TOV_Gen} and using the TOV equations for the isotropic solution, we find
\begin{equation}	
	\begin{aligned}
	\label{1p1p2_definitions_eq:TOV_Ans_K}
		\mathcal{K}_{,\rho} & =-2\mathcal{K}\left(\mathcal{K}-\mathbb{M}_\ast \frac{\mathcal{K}}{\mathcal{K}_\ast} - \bar{\mathbb{M}}-\frac{1}{4}\right)\,,
	\end{aligned}
\end{equation}
and the algebraic equations
\begin{equation}
	\begin{aligned}
		\label{1p1p2_definitions_eq:TOV_Ans}
		\mathds{P}& = \frac{1}{6} \bar{\mathbb{M}} \left(1+4P_\ast-4\mathcal{K}\right) - 2P_\ast \left(\mathcal{K}-\mathcal{K}_\ast\right)\\
		& +\frac{\mathbb{M}_\ast}{6\mathcal{K}_\ast} \left(\mathcal{K}-\mathcal{K}_\ast\right) \left(4P_\ast+1-4\mathcal{K}_\ast-4\mathcal{K}\right)\,,\\
		E & = \frac{1}{3} \left(\mathbb{M}_\ast \frac{\mathcal{K}}{\mathcal{K}_\ast} + \bar{\mathbb{M}}\right) -\frac{1}{2}\mathds{P}-\mathcal{K}+\frac{1}{4}\,,\\
		X & = \mathcal{K}-\mathbb{M}_\ast \frac{\mathcal{K}}{\mathcal{K}_\ast} - \bar{\mathbb{M}}-\frac{3}{4}\,,\\
		Y & = \mathcal{K}+P_\ast-\frac{1}{4}\,,\\
		\bar{\mathbb{M}} & = K_0 e^\rho \mathcal{K} \left(\mu-\mu_\ast\right)\,.
	\end{aligned}
\end{equation}

Equations~\eqref{1p1p2_definitions_eq:Ansatz_M_K}--\eqref{1p1p2_definitions_eq:TOV_Ans} characterize a type of generating theorem not considered in \cite{Rahman,Lake, Martin,Boonserm:2005ni,Boonserm:2006up,ExactSolutions}, where the deformations that relate the new quantities and those associated with the isotropic solution are not independent. This type of mixing allows us to reduce the system of differential equations to a single differential equation for $\mathcal{K}$. The remaining variables are solutions of algebraic equations.
Moreover, Eq.~\eqref{1p1p2_definitions_eq:TOV_Ans_K} is a Bernoulli differential equation. Given a particular isotropic spacetime and specifying the function $\bar{\mathbb{M}}$, Eq.~\eqref{1p1p2_definitions_eq:TOV_Ans_K} can be formally solved.

In the characterization of the new solutions, the covariant 1+1+2 potentials in Eq.~\eqref{1p1p2_definitions_eq:Normalized_vars} fully characterize the geometry and thermodynamics. Nonetheless, once these quantities are calculated, we can break covariance to write the solution in the more traditional form in terms of the components of the metric tensor and the thermodynamic variables of the fluid source. For instance, if the 2-surfaces are isotropic submanifolds, the circumferential radius, $r$, and the parameter $\rho$ are related by
\begin{equation}
	\label{1p1p2_definitions_eq:rho_r_relation}
	\rho = 2 \ln \left(\frac{r}{r_0}\right)\,.\
\end{equation}
where $r_0^2=K_0$. Then, considering a spacetime characterized by a line element of the form
\begin{equation}
	\label{1p1p2_definitions_eq:general_line_element}
	ds^2 = -g_{tt}(r)~dt^2 + g_{rr}(r)~dr^2 + r^2 d\Omega^2\,,
\end{equation}
where $d\Omega^2$   represents the line element of the unit 2-sphere, we have
\begin{equation}
	\label{1p1p2_definitions_eq:variables_metric_relation}
	\begin{aligned}
		\mathcal{K} &=  \frac{g_{rr}}{4},\qquad
		&Y = \frac{r}{4 g_{tt}}\frac{d g_{tt}}{dr}\,,\qquad
			&\phi &=\frac{2}{r\sqrt{g_{rr}}}\,,
	\end{aligned}
\end{equation}
whereas the thermodynamic variables are obtained as
\begin{equation}
	\begin{aligned}
		\mu &= \mathbb{M}\phi^2\,, \qquad
		&p_r &=\left(P+\mathds{P}\right)\phi^2\,, \qquad
		&p_t &=\left(P-\frac{1}{2}\mathds{P}\right)\phi^2\,.
	\end{aligned}
\end{equation}

\section{Generating theorems and the uniqueness of interior solutions}\label{Unique_Sec}
Generating theorems can be employed to derive and investigate some properties of interior solutions. Here, we will use the theorems described in the previous section to show that the interior Schwarzschild solution is unique from the point of view of the generating theorems, in the sense that any other solution with constant density is not physically acceptable. We will also show an (unsuccessful) attempt to generate a constant density anisotropic interior solution from the interior Schwarzschild one. This negative result will motivate those reported in the next sections.

Consider the Interior Schwarzschild solution, characterized by the line element, in Schwarzschild coordinates,
\begin{equation}
	\label{linelement_IS_G0}
	d s^2 = -A_0\left(\sqrt{3 - \mu _\ast r^2}+P_0\right)^2 d t^2+ \frac{3}{3 -\mu _\ast r^2} d r^2 +r^2\,\big(d\theta^2+\sin^2\!\theta d\phi^2\big)\,,
\end{equation}
where $A_0$ and $P_0$ are integration constants, with a fluid source characterized by a constant energy density $\mu_\ast$ and pressure
\begin{equation}
p=-\frac{\mu _\ast \left(P_0+3 \sqrt{3-\mu _\ast r^2}\right)}{3 \left(P_0+\sqrt{3-\mu _\ast r^2}\right)}\,.
\end{equation}
In terms of the covariant variables, this solution is fully characterized by
\begin{equation}
\begin{split}
	\label{IS_1p1p2_potentials}
\mathcal{K}_\ast &=\frac{3}{12 -4 \mu _\ast K_0 e^\rho }\,,\\
\mathbb{M}_\ast &=\mu _\ast K_0e^\rho \mathcal{K}_\ast\,, \\
P_\ast &=\frac{\mu _\ast K_0e^{\rho } \left(3 \sqrt{3-\mu _\ast K_0e^{\rho}}+P_0\right)}{4 \left(\mu_\ast K_0e^{\rho }-3\right) \left(\sqrt{3-\mu _\ast K_0e^{\rho }}+P_0\right)}\,.
\end{split}
\end{equation}
Using the transformations \eqref{PY-theorem} we have
\begin{equation}
\bar{P}_{,\rho }+\bar{P}^2+\frac{\sqrt{3-\mu _\ast K_0e^{\rho }} \left(\mu _\ast K_0e^{\rho }+6\right)+P_0\left(6-\mu _\ast K_0e^{\rho }\right)}{2 \left(\mu _\ast K_0e^{\rho }-3 \right)
   \left(\sqrt{3-\mu _\ast K_0e^{\rho }}+P_0\right)}\bar{P}=0
\end{equation}
which has the solution
\begin{equation}
	\bar{P}=-\frac{\mu _\ast K_0e^{\rho }}{\sqrt{3-\mu _\ast K_0e^{\rho }}} \left[  P_0 \left(2-P_0 \bar{P}_0\right)-2 \left(1-P_0\bar{P}_0\right) \sqrt{3-\mu _\ast K_0e^{\rho }}+\bar{P}_0 \left(3-K_0 \mu_0 e^{\rho }\right)\right]^{-1}\,,
\end{equation}
where $\bar{P}_0 $ is an integration constant. This solution yields the line element
\begin{equation}
d s^2 = -A_0 \left[\bar{P}_0 \left(P_0+\sqrt{3-\mu _\ast r^2}\right)-2\right]^2 d t^2+ \frac{3}{3 -\mu _\ast r^2} d r^2 +r^2\,\big(d\theta^2+\sin^2\!\theta d\phi^2\big)\,,
\end{equation}
and leads to an isotropic pressure profile of the fluid,
 \begin{equation}
 	\label{linelement_IS_G1}
 p=\frac{\mu _*}{3}  \left[\frac{2 \bar{P}_0 \sqrt{3-\mu _* r^2}}{2-\bar{P}_0 \left(P_0+\sqrt{3-\mu _* r^2}\right)}-1\right]
 \end{equation}
It is not difficult to prove that, although superficially different, the new solutions and the interior Schwarzschild solution differ only by the value of the integration constants. That is, we recover the original spacetime and no additional constant density solution can be generated by the $PY$-theorem.

This result is somewhat expected. Since the TOV equations are ordinary differential equations, the only freedom compatible with the generating theorems is the equation of state: in the transformation, the equation of state of the initial solution is modified, and this brings a modification of the metric.  In the constant density case, the TOV equations do not have that freedom, and the unicity theorem requires that the interior Schwarzschild solution will be mapped onto itself. Incidentally, this also implies that the interior Schwarzschild solution is unique, in the sense that no other static, isotropic, constant-density solution of the Einstein field equations, with vanishing cosmological constant, exists.

Instead of applying the $PY$ theorem, one might try to apply the $\mathbb{P}Y$ theorem to the interior Schwarzschild solution to generate a solution with non-trivial anisotropic pressure. In that case, Eq.~\eqref{PbbY-theorem} reads
\begin{equation}
	\label{EqPPGen1}
\bar{\mathds{P}}_{,\rho }+\bar{\mathds{P}}^2+\frac{\sqrt{3-\mu _\ast K_0  e^\rho } \left(4 \mu_\ast K_0  e^\rho -3\right)+P_0 \left(2 \mu _\ast K_0  e^\rho -3\right)}{2 \left(\mu _\ast K_0 e^\rho -3\right) \left(\sqrt{3-\mu _\ast K_0  e^\rho }+P_0\right)}\bar{\mathds{P}}=0\,,
\end{equation}
which has the solution
\begin{equation}
\begin{aligned}
\bar{\mathds{P}}= & 3 \left(3-P_*^2\right)^{5/2}\left\{2\sqrt{3-P_0^2}\sqrt{3-\mu _\ast K_0 e^{\rho }} \left[ P_0 \left(2 P_0^2+3\right)\mu _\ast K_0 e^{\rho }\right.  \right.\\
&\left.+\sqrt{3-\mu _\ast K_0 e^{\rho }} \left(\left(P_0^2+6\right) \mu _\ast K_0 e^{\rho}-\left(P_0^2-3\right)^2\right)\right] \\
&\left. -9 P_0 \sqrt{\mu _\ast K_0 e^{\rho} }  \left(\sqrt{3-\mu _\ast K_0 e^{\rho}}+P_0\right)^2 \tanh^{-1}\left(\frac{ \sqrt{(3-P_0^2)\mu _\ast K_0 e^{\rho}}}{P_0 \sqrt{3-\mu _\ast K_0 e^{\rho }}+3}\right)\right\}^{-1}\,.
\end{aligned}
\end{equation}
Unlike the case of the $PY$ theorem, it is not possible in this case to present the solution for the metric tensor using the radial coordinate, as it is required to perform a further integration. Nonetheless, it is straightforward to calculate the expression for the radial pressure, such that
\begin{equation}
\begin{split}
p_r=&-\frac{\mu _0 \left(P_0+3 \sqrt{3-\mu _0 r^2}\right)}{3 \left(P_0+\sqrt{3-\mu _0 r^2}\right)}+\\
& +2 \left(3-P_0^2\right){}^{5/2} \sqrt{3-\mu _* r^2}\left\{2 r^2\sqrt{3-P_0^2}\sqrt{3-\mu _\ast r^2} \left[ P_0 \left(2 P_0^2+3\right)\mu _\ast r^2\right.  \right.\\
&\left.+\sqrt{3-\mu _\ast r^2} \left(\left(P_0^2+6\right) \mu _\ast r^2-\left(P_0^2-3\right)^2\right)\right] \\
&\left. -9  P_0 r^3 \sqrt{\mu _\ast  }   \left(\sqrt{3-\mu _\ast r^2}+P_0\right)^2 \tanh^{-1}\left(\frac{ \sqrt{(3-P_0^2)\mu _\ast r^2}}{P_0 \sqrt{3-\mu _\ast r^2}+3}\right)\right\}^{-1}\,.
\end{split}
\end{equation}
which leads to a divergent  anisotropic pressure $\Pi$ at $r=0$ and negative in an open neighborhood of $r=0$. These features makes this solution unphysical. Applying the $\mathbb{P}Y$ theorem again could, in principle, produce physical solutions. However, the procedure yields a differential equation with transcendental coefficients that is not easy to integrate.

These results show that it is not immediate to generate an analytic, closed-form, anisotropic, constant density physical solution of the TOV equations by means of the $\mathbb{P}Y$ theorem. However, as we will see, other generating theorems can lead to closed-form, physically interesting solutions.

\section{ Generation of new anisotropic solutions}\label{NewSol_Sec}
\subsection{Constant density solution}

The new $P\mathds{P}$ generating theorem specifically maps solutions of the TOV equation whose fluid sources are not necessarily of the same type, allowing the generation of solutions with non-trivial anisotropic stresses from solutions with isotropic fluid sources.


As in the previous section, consider as a base isotropic solution the interior Schwarzschild spacetime, characterized by a constant energy density, $\mu_\ast$, a line element~\eqref{linelement_IS_G0}, and the 1+1+2 potentials in Eq.~\eqref{IS_1p1p2_potentials}.
Let the new spacetime also be a solution of the Einstein field equations for a constant density fluid source, $\mu$, such that $\mu= \mu_\ast + \bar{\mu}= \text{constant}$.

Using the results from Section~\ref{Sec:1+1+2TOVeq}, we find
\begin{equation}
	\begin{aligned}
		\label{new_solutions_eq:Const_density_NewSol_thermodynamic_variables_cov}
		\mathcal{K} & =\frac{3}{12- 4 K_0 e^\rho \mu}\,,\\
		\mathbb{M} & =\frac{3 K_0 e^\rho \mu}{12- 4 K_0 e^\rho \mu}\,,\\
		P & = \frac{K_0 e^\rho \mu_\ast \left( P_0 + 3\sqrt{3 -K_0 e^\rho \mu_\ast} \right)}{4\left(K_0 e^\rho \mu_\ast -3 \right) \left(P_0 + \sqrt{3- K_0 e^\rho \mu_\ast} \right)}\\
		&-\frac{3K_0^2e^{2\rho}\left(\mu-\mu_*\right)\left[\mu\left(2 K_0e^{\rho}\mu_*-3\right)-3\mu_*\right]}{8\left(K_0e^{\rho}\mu-3\right)^2\left(K_0e^{\rho}\mu_*-3\right)^2}\,,\\
		\mathds{P} & = \frac{3K_0^2e^{2\rho}\left(\mu-\mu_*\right)\left[\mu\left(2 K_0e^{\rho}\mu_*-3\right)-3\mu_*\right]}{8\left(K_0e^{\rho}\mu-3\right)^2\left(K_0e^{\rho}\mu_*-3\right)^2}\,.        
	\end{aligned}
\end{equation}

From Eqs.~\eqref{1p1p2_definitions_eq:rho_r_relation} and \eqref{1p1p2_definitions_eq:variables_metric_relation}, we find the new solution is characterized by a line element of the form~\eqref{1p1p2_definitions_eq:general_line_element} with the non-trivial metric coefficients
\begin{equation}
	\begin{aligned}
		\label{new_solutions_eq:Const_density_NewSol_metric}
		g_{tt} & = \frac{A \sqrt{3-\mu_\ast r^2} \left(P_0 + \sqrt{3-\mu_\ast r^2}\right)^2}{3\sqrt{3-\mu r^2}}\,, \qquad
		& g_{rr} & = \frac{3}{3 -\mu r^2}\,,
	\end{aligned}
\end{equation}
where $A$ is an integration constant. Naturally, for $\mu=\mu_\ast$ we recover the interior Schwarzschild solution.

\subsection{Non-constant density solutions}

The solution found in the previous subsection follows from deforming the interior Schwarzschild spacetime and imposing that a constant-density fluid permeates the resulting spacetime.  Using the $P\mathds{P}$ theorem, we may consider more general deformations. For instance, consider a deformation of the form
\begin{equation}
	\label{new_solutions_eq:Non_Const_density_general_form}
	\mu=\mu_\ast+\alpha+\sum_{n=1}^{N}c_{n}K_{0}^{\frac{n}{2}}e^{\frac{n}{2}\rho}\,,
\end{equation}
where $N\in\mathbb{N}_{>0}$, and $\alpha$, $c_i$, with $i=2,3,...$, are constants.
This type of deformation may be used to generate a wide class of solutions with non-trivial anisotropic pressure. In particular, $\mu$ may be an analytic function, such that the sum in Eq.~\eqref{new_solutions_eq:Non_Const_density_general_form} is regarded as a power series in $r$.

Using Eq.~\eqref{new_solutions_eq:Non_Const_density_general_form}, Eqs.~\eqref{1p1p2_definitions_eq:TOV_Ans_K} can be readily integrated, finding
\begin{equation}
	\label{new_solutions_eq:Non_Const_density_K}
	\mathcal{K} = \frac{1}{4\left[1-\frac{1}{3} \left(\mu_\ast+\alpha\right) K_0 e^\rho - \sum_{n=1}^N\frac{c_n}{n+3} K_0^{\frac{n+2}{2}} e^{\frac{n+2}{2}\rho}\right]}\,.
\end{equation}
then, using this result in Eq.~\eqref{1p1p2_definitions_eq:TOV_Ans} we can find closed form expressions for the remaining 1+1+2 potentials. This case clearly showcases an advantage of the covariant approach. Using Eq.~\eqref{new_solutions_eq:Non_Const_density_K} in
Eq.~\eqref{1p1p2_definitions_eq:TOV_Ans}, yields, in particular,
\begin{equation}
	\label{new_solutions_eq:Non_Const_density_Y}
	\begin{aligned}
		Y & = \frac{1}{4\left[1-\frac{1}{3} \left(\mu_\ast + \alpha\right) K_0 e^\rho - \sum_{n=1}^N\frac{c_n}{n+3} K_0^{\frac{n+2}{2}} e^{\frac{n+2}{2}\rho}\right]}\\
		& - \frac{3K_0 e^\rho \mu_\ast \left(\sqrt{3- \mu_\ast K_0 e^\rho }-P_0\right)}{4\left(3-K_0 e^\rho \mu_\ast \right) \left(\sqrt{3- \mu_\ast K_0 e^\rho} - 3P_0\right) }- \frac{1}{4}\,.
	\end{aligned}
\end{equation}

The 1+1+2 potentials fully characterize the geometry of the new solution and the properties of the matter fluid, containing the same information as the metric tensor. In addition, using Eq.~\eqref{1p1p2_definitions_eq:variables_metric_relation}, we can formally find the non-trivial metric coefficients, provided the integrals can be solved analytically. One such case follows from setting $N=2$, $c_1=0$ and $c_2=\beta$, such that, 
\begin{equation}
	\mu=\mu_\ast + \alpha + \beta K_0 e^\rho=\mu_\ast + \alpha + \beta r^2\,.
\end{equation}
In this case, the new solution is characterized by the line element~\eqref{1p1p2_definitions_eq:general_line_element} with
\begin{equation}\label{Sol_Var_mu}
	\begin{aligned}
		g_{tt} & = A \frac{\sqrt{3-\mu_\ast r^2} \left(P_0+\sqrt{3-\mu_\ast r^2}\right)^2}{\sqrt[4]{15-5\left(\alpha+\mu_\ast\right) r^2 -3\beta r^4}} \left[\frac{5\left(\alpha+\mu_\ast\right) +4\Gamma\left[5\left(\alpha+\mu_\ast\right) +6\beta r^2\right]}{5\left(\alpha+\mu_\ast\right) -4\Gamma\left[5\left(\alpha+\mu_\ast\right) +6\beta r^2\right]}\right]^{\Gamma}\,,\\
		g_{rr} & = \frac{15}{15-5\left(\alpha+\mu_\ast\right)r^{2}-3 \beta r^4}\,,
	\end{aligned}
\end{equation}
where $A$ is a constant and
\begin{equation}
	\Gamma = \frac{\sqrt{5}\left(\alpha+\mu_\ast\right)}
	{4\sqrt{5\left(\alpha+\mu_\ast\right)^2 +36\beta}}\,.
\end{equation}
The pressures are given by
\begin{align*}
p_r & = \frac{\mu_{\ast}\left(P_0+3\sqrt{3-\mu_{\ast}r^2}\right)\left[3\beta r^4+5r^2 \left(\alpha+\mu_{\ast}\right)-15\right]}{15\left(3-\mu_{\ast}r^2\right) \left(P_0+\sqrt{3-\mu_{\ast}r^2}\right)}\,,\\
p_{t}-p_{r} & =\frac{r^2 \left(15 \alpha +15\beta  r^2-2 \beta  \mu _* r^4\right) \left\{6 \beta \mu _* r^4+r^2 \left[10 \mu _* \left(\alpha +\mu _*\right)-9 \beta \right]-15 \left(\alpha +2 \mu _*\right)\right\}}{20 \left(3-\mu _* r^2\right)^2 \left[3\beta  r^4+5 r^2 \left(\alpha +\mu _*\right)-15\right]}\\
&+\frac{\beta  \mu _* r^4}{5\sqrt{3-\mu _* r^2} \left(P_0+ \sqrt{3-\mu _* r^2}\right)}\,.
\end{align*}

\section{Properties of the new solutions}\label{Sec:NewSol_Prop}
\subsection{Regularity and classical energy conditions for the constant density solution}
In this section, we will study the features of the new solution~\eqref{new_solutions_eq:Const_density_NewSol_metric} in detail. In that regard, it is convenient to express the non-trivial metric coefficients in the following form
\begin{equation}
	\label{Constant_density_sol_eq:metric_rewritten}
	\begin{aligned}
		g_{tt} & =\frac{A \sqrt{1-B \frac{r^2}{r_b^2}} \left(\sqrt{1-B\frac{r^2}{r_b^2}} -3 \sqrt{1-B}\right)^2}{\sqrt{1-\frac{2M r^2}{r_b^3}}}\,,
		\\
	g_{rr} & =\left(1-\frac{2Mr^{2}}{r_{b}^{3}}\right)^{-1}\,,
	\end{aligned}
\end{equation}
namely, comparing with Eq.~\eqref{new_solutions_eq:Const_density_NewSol_metric}, we have set
\begin{equation}
	\begin{aligned}
		B & =\frac{1}{3}r_b^2\mu_{\ast}\,,\qquad
		& M & =\frac{1}{6}r_{b}^{3}\mu\,,\qquad
		&P_0 & =-3 \sqrt{3-3B}\,,
	\end{aligned}
\end{equation}
and $r_b>0$ is an otherwise arbitrary constant. The choice of $P_0$ is such that at $r=r_b$ the solution can be smoothly matched to a branch of the vacuum Schwarzschild solution. In that case, $M$ represents the gravitational mass parameter characterizing the fluid source, such that
\begin{equation}
	M = \frac{1}{2} \int_0^{r_b} \mu\,x^2 dx\,.
\end{equation}
Then, the new solution can be used to characterize the interior of a static, spherically symmetric compact stellar object with circumferential radius  $r_b$, such that $r\in \left[0,r_b\right]$.

In the form~\eqref{Constant_density_sol_eq:metric_rewritten}, it is clear that
the geometry of the spacetime is completely characterized by the parameter $B$ and $M/r_b$. The latter is known as the compactness parameter.

In what follows, it is important to analyze the behavior of the radial and tangential pressures. Using, Eq.~\eqref{new_solutions_eq:Const_density_NewSol_thermodynamic_variables_cov} we find
\begin{equation}
	\begin{aligned}
		\label{Constant_density_sol_eq:pressures}
		p_r & = \frac{3B\left(\sqrt{1-B}-\sqrt{1-B\frac{r^{2}}{r_b^2}}\right)\left(1-2M\frac{r^2}{r_b^3}\right)}{r_b^2\left(3\sqrt{1-B}-\sqrt{1-\frac{Br^2}{r_b^2}}\right)\left(B\frac{r^2}{r_b^2}-1\right)}\,,\\
		p_t-p_r & = \frac{3r^2\left(B-\frac{2M}{r_b}\right)\left(4BM\frac{r^2}{r_b^3}-B-\frac{2M}{r_b}\right)}{4r_b^4\left(1-B\frac{r^2}{r_b^2}\right)^2\left(1-2M\frac{r^2}{r_b^3}\right)}\,.
	\end{aligned}
\end{equation}

Considering Eqs.~\eqref{Constant_density_sol_eq:pressures}, the radial and tangential pressures will diverge at the characteristic radii
\begin{equation}
	\begin{aligned}
		r_1 &=\frac{r_{b}}{\sqrt{B}}\,,\\
		r_2 &= r_b \sqrt{\frac{r_b}{2M}}\,,\\
		r_3 &= r_b\sqrt{9-\frac{8}{B}}\,.
	\end{aligned}
\end{equation}
Computing the Ricci scalar, we see that at these radii the solution is singular. Therefore, to have a regular solution for all $r \leq r_b$ we must set $B < \frac{8}{9}$ and $r_b>2M$. The latter constraint is expected: the radius of the star must be greater than its Schwarzschild radius.

\begin{figure}[h]
	\centering
	\includegraphics[width=1\columnwidth]{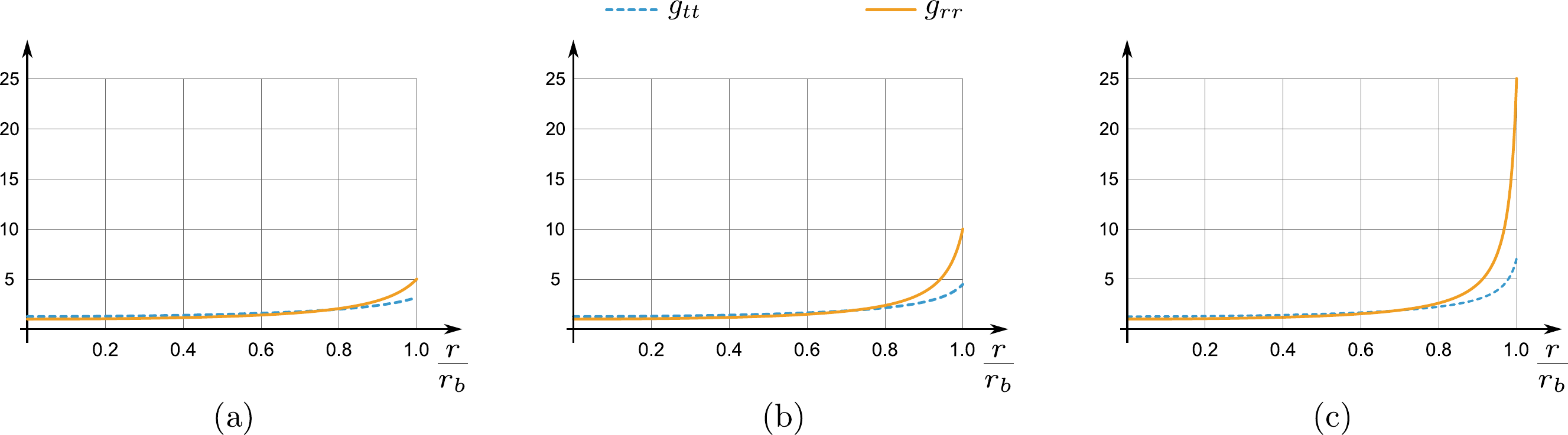}
	\caption{
		\label{Figure:Generalized_In_Schw_metric_plots}
		Non-trivial metric components, for various
		values of the compactness parameter, $M/r_b$, for the 
		solution~\eqref{Constant_density_sol_eq:metric_rewritten}.
		In all cases, we have set $A=1$ and $B=0.5$. In (a) $M/r_b=0.4$,
		(b) $M/r_b=0.45$ and (c) $M/r_b=0.48$.
	}
\end{figure}

In Figure~\ref{Figure:Generalized_In_Schw_metric_plots}, we present the
behavior of the $g_{tt}$ and $g_{rr}$ metric components for various values of the 
parameter $M/r_b$ and fixed $B$. Note that the $g_{rr}$ component grows very fast as $r$ approaches $r_b$, however its graph does
not have an asymptote at $r=r_b$, unless $r_b=2M$.

\begin{figure}[h]
\centering
\includegraphics[width=1\columnwidth]{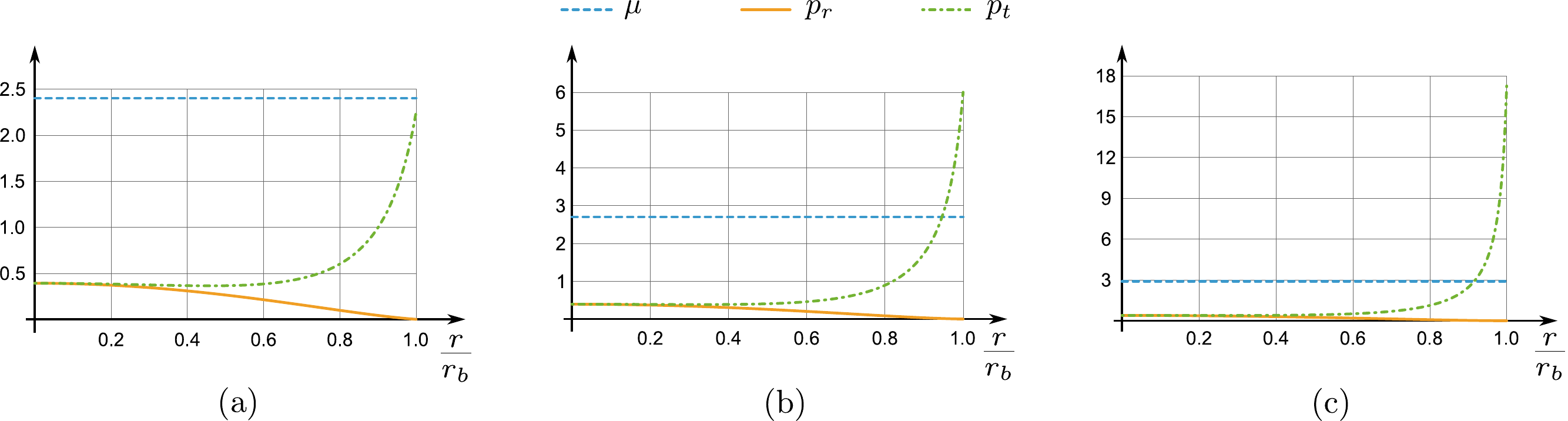}
\caption{
	\label{Figure:Generalized_In_Schw_matter_plots}
	Energy density, radial and
	tangential pressures as functions of the radial coordinate, for various
	values of the compactness parameter, $M/r_b$, for the 
	solution~\eqref{Constant_density_sol_eq:metric_rewritten}.
	In all cases, we have set $A=1$ and $B=0.5$. In (a) $M/r_b=0.4$,
	(b) $M/r_b=0.45$ and (c) $M/r_b=0.48$.
	}
\end{figure}

For $B < \frac{8}{9}$ and $r_b>2M$, $p_r$ is always non-negative and vanishes at $r=r_b$, only. The tangential pressure, $p_t$, may
take negative values depending on the relative values of $M/r_b$ and $B$. Below we present sufficient conditions on the value of $B$, depending on the values of $M/r_b$, for the tangential pressure to remain positive for all $r \in \left[0,r_b\right]$:
\begin{equation*}
	\left\{ 
		\begin{aligned} 
			& \text{If }M=0\text{, then }B=0\\
			& \text{If }0<\frac{M}{r_{b}}<\frac{4}{9}\text{, then } 0\leq B\leq\frac{2M}{r_{b}}\\
			& \text{If }\frac{4}{9} \leq \frac{M}{r_{b}}<\frac{1}{2}\text{, then }0 \leq B < \frac{8}{9}
		\end{aligned}
	\right.
\end{equation*}
Figure~\ref{Figure:Generalized_In_Schw_matter_plots} shows the
behavior of the matter variables for various values of the 
parameter $M/r_b$ and fixed $B$. Similarly to the behavior of the $g_{rr}$ component of the metric, the tangential pressure grows very fast as $r$ approaches $r_b$, but its graph does not have an asymptote at $r=r_b$, unless $r_b=2M$.

Although an important property of the fluid source of the new solution, it is not necessary to require that the pressures remain positive. Indeed, to support the interpretation of the solution in Eq.~\eqref{Constant_density_sol_eq:metric_rewritten} as a model for the interior of an astrophysical compact stellar object, we may only require that the source fluid satisfies the pointwise weak and strong energy conditions, for all $r\in \left[0, r_b\right]$.

The weak energy condition requires that
\begin{equation}
	\label{Constant_density_sol_eq:WEC}
	\left\{
	\begin{aligned}
		\mu & \geq0\\
		\mu+p_r & \geq0\\
		\mu+p_t & \geq0
	\end{aligned}
	\right.
\end{equation}
whereas the strong energy condition requires that
\begin{equation}
	\label{Constant_density_sol_eq:SEC}
	\left\{
	\begin{aligned}
		\mu+p_r & \geq0\\
		\mu+p_t & \geq0\\
		\mu+p_r+2 p_t & \geq0
	\end{aligned}
	\right.
\end{equation}

The values of $B$ for which the energy conditions are verified depend strongly on $M/r_b$, and the expressions for the conditions on $M/r_b$ for which the weak or the strong energy condition are verified are rather large, so we will not list them here. However, for any value of the compactness parameter, from zero to the black hole limit, the new solution is regular, and there is a range of values of $B$ for which the fluid source verifies the weak and strong energy conditions throughout the star.

\subsection{Relation of the constant density solution with other solutions}
Constant density solutions of the TOV equations, namely the Interior Schwarzschild, the Florides~\cite{Florides1974abc} and the Bowers-Liang~\cite{Bowers_Liang_1974} solutions, have been extensively considered in the literature as limiting cases of physical configurations (see, e.g.,~\cite{Yagi_Yunes_2015,Silva_et_al_2015,Singh_et_aj_2019,Olmo_et_al_2020,Becerra_2026} and references therein). The former two are connected by the new constant density solution found here.

As we have discussed above, in the particular case of $B=\frac{2M}{r_b}$,
Eqs.~\eqref{Constant_density_sol_eq:pressures} imply $p_r=p_t$, and
Eq.~\eqref{Constant_density_sol_eq:metric_rewritten}  reduces to the metric
components of the Interior Schwarzschild solution. In this sense, the new constant
density solution directly generalizes the Interior Schwarzschild spacetime, and we 
can recover the metric of the Interior Schwarzschild spacetime continuously.  This, 
however, also implies that, as for the latter solution, a spacetime characterized by 
Eq.~\eqref{Constant_density_sol_eq:metric_rewritten} is, strictly speaking, 
nonphysical, since it would characterize a star with a homogeneous energy density. 
Nonetheless, as for the Interior Schwarzschild solution, it can be considered as a 
limiting case for compact stellar objects with non-trivial shearing stresses.

For $B=0$, the radial pressure, Eq.~\eqref{Constant_density_sol_eq:pressures}, is identically zero. In that exotic case, the interior spacetime can be thought of as being permeated by
a fluid composed of concentrically infinitesimally thin matter shells
with no direct thermodynamic interaction between them, and entirely
held against collapse by tangential stresses. It is trivial to show
that for $B=0$ the metric of the spacetime reduces to the unique, spherically symmetric, non-trivial solution of the Einstein field equations with
constant energy density, identically zero radial pressure, and zero
cosmological constant: the Florides solution~\cite{Florides1974abc}.
These results suggest, from a mathematical point of view, that the Interior Schwarzschild and the Florides solutions are not isolated solutions of the Einstein field equations.

Another notable case is the configuration that follows from considering $M=0$. In this case, Eq.~\eqref{Constant_density_sol_eq:metric_rewritten} does not necessarily reduce to the Minkowski metric. Indeed, if $B\neq 0$,  the energy density is also zero throughout the compact stellar object. However, pressures are present: the radial pressure is always non-negative, but the tangential pressure is positive in a neighborhood around the center and negative in some region toward the outer boundary.

Solutions with vanishing gravitational mass have been dubbed {\it ghost stars} in~\citep{Zeldovich_Novikov_Book_1971,Herrera_Prisco_Ospino_2024}. However, previous models for these objects are singular at the center ($r=0$) and exhibit regions where the energy density is negative. The solution presented here has the same structure, but avoids these problems: for $M=0$ and $B<\frac{8}{9}$ there are no curvature singularities for $r\leq r_b$, and the energy density is identically zero throughout the star.
It should be noted, however,  that in a region toward the boundary, at $r=r_b$, the null energy condition is always violated. Nonetheless, as this violation does not occur  throughout the
star, this allows for the possibility of matching the interior region, with zero energy density, with an
exterior spacetime through a thin matter shell at the matching surface, at some $r<r_b$, such that the weak and the strong energy conditions are verified throughout the spacetime.

On the other hand, if we allow the classical energy conditions to be violated, the zero mass parameter solution presents the possibility of a curious setup. By construction, the new solution is characterized by vanishing radial pressure, $p_r$, at $r=r_b$. Then, it is possible to smoothly match the ghost star solutions at $r=r_b$ with an exterior Minkowski spacetime. 
This exotic object would then represent a region of the spacetime with non-trivial geometry, changing the paths of particles that go through it, but undetectable by measuring the paths of gravitational masses strictly outside this region.

Finally, it is also useful to list, for comparison, the properties of the new solution with those of the best-known anisotropic solution with constant density, the Bowers-Liang solution, characterized by the line element~\eqref{1p1p2_definitions_eq:general_line_element} with
 \begin{equation}
	\begin{aligned}
		\label{BLSol}
	g_{tt} &=A_0 \left[\left(1-\frac{2M}{r_{b}^{3}}r^{2}\right)^{h/2}-3\left(1-\frac{2M}{r_{b}}\right)^{h/2}\right]^{2/h}
	\,, \\
	g_{rr} & =
	 \left(1-\frac{2M}{r_{b}^{3}}r^{2}\right)^{-1}\,,
	\end{aligned}
\end{equation}
where $A_0 \in \mathbb{R}_{>0}$, $h\in \mathbb{R}_{\neq0}$, and $r_b > 2M$. For $h=1$, we see that the Bowers-Liang reduces to the Interior Schwarzschild solution.

For completeness, we present below the expressions for the pressures of fluid source of the Bowers-Liang spacetime:
\begin{equation}
	\begin{aligned}
		p_r & = -\frac{6M}{r_b^3}
		\frac{\left(1-\frac{2Mr^2}{r_b^3}\right)^{h/2}-\left(1-\frac{2M}{r_b}\right)^{h/2}}{\left(1-\frac{2Mr^2}{r_b^3}\right)^{h/2}-3\left(1-\frac{2M}{r_b}\right)^{h/2}}\,,\\
		p_t-p_r & = \frac{12\left(1-h\right)M^2r^2\left(1-\frac{2M}{r_b}\right)^{h/2}\left(1-\frac{2Mr^2}{r_b^3}\right)^{h/2}}{r_b^6\left(1-\frac{2M}{r_b^3}r^2\right)\left[\left(1-\frac{2Mr^2}{r_b^3}\right)^{h/2}-3\left(1-\frac{2M}{r_b}\right)^{h/2}\right]^2}\,,
	\end{aligned}
\end{equation}

The Bowers-Liang solution has a number of interesting properties, however, its nontrivial dependence on the parameter $h$ makes it difficult to analyze. Nonetheless, it is possible to prove some general features of this solution. For instance, for any values of the parameters, the radial pressure vanishes at $r=r_b$, whereas the tangential pressure is negative at that surface if $h>1$, zero if $h=1$, and positive if $h<1$. On the other hand, it is straightforward to verify that if the energy density is positive, the parameter $h$ is positive and
\begin{equation}
	\frac{2M}{r_{b}} \geq 1-\left(\frac{1}{9}\right)^{\frac{1}{h}}\,,
\end{equation}
the radial and the tangential pressures of the fluid source of the Bowers-Liang solution will diverge at some $r\in \left[ 0,r_b\right]$. That is, for $h>0$, the Bowers-Liang solution cannot have a compactness parameter arbitrary close to the black hole limit: $2M/r_b=1$, without the existence of curvature singularities.

For $h<0$, the solution is regular for any value of the compactness parameter until the black hole limit of $2M/r_b=1$. For $M \geq 0$, the weak and the strong energy conditions are verified throughout the stellar compact object. Nonetheless, the fluid source will be characterized by radial tension for all $r\in \left[0,r_b\right]$, and tangential tension in a region around the core of the star.

These issues suggest that the use of the Bowers-Liang solution might be less advantageous than the use of solution \eqref{Constant_density_sol_eq:metric_rewritten} to model the maximally compact state of an anisotropic compact stellar object.

\subsection{Some properties of the variable density solution}

We will now focus on the solution with variable density \eqref{Sol_Var_mu}. As in the constant-density case, it is convenient to reparameterize the metric. Defining, 
\begin{equation}
M_1=\frac{r_b^5}{5}\beta,\quad M_2=\frac{r_b^3}{3}(\alpha+\mu_\ast),\quad B=\frac{r_b^2}{3}\mu_\ast, \quad P_0=-3\sqrt{3-3B}\,,
\end{equation}
the total mass is $2M=M_1+M_2$, and the compactness is given by the ratio $M/r_b$.  Using these parameters, the metric coefficients read
\begin{equation}
	\label{Sol_Var_mu_Par}
	\begin{aligned}
		g_{tt} & = A_0 \frac{\sqrt{1-B\frac{r^2}{r_b^2}}}{\left(1-\frac{M_2 r^2}{r_b^3}-\frac{M_1 r^4}{r_b^5}\right)^{1/4}} \left(\sqrt{1-B\frac{ r^2}{r_b^2}}-3 \sqrt{1-B}\right)^2
& \left[\frac{2 M_2 }{(1-4 \Gamma ) M_2- 8 \Gamma  M_1 \frac{ r^2}{r_b^2}}-1\right]^{\Gamma}\,,
\\
g_{rr} & = \left(1-\frac{M_2 r^2}{r_b^3}-\frac{M_1 r^4}{r_b^5}\right)^{-1}\,,
	\end{aligned}
\end{equation}
where $A_0 \in \mathbb{R}$ and
\begin{equation}
	\Gamma = \frac{M_2}{4 \sqrt{4 M_1 r_b+M_2^2}}\,.
\end{equation}
In the new variables, the pressures are given by
\begin{equation}
	\begin{aligned}
		p_r  & = \frac{3B\left(\sqrt{1-B}-\sqrt{1-\frac{Br^2}{r^2_b}}\right) \left(1-M_1\frac{r^4}{r^5_b}-M_2\frac{r^2}{r^3_b}\right)}{r^2_b \left(1-B\frac{r^2}{r^2_b}\right) \left(\sqrt{1-B\frac{r^2}{r^2_b}}-3\sqrt{1-B}\right)}\,,\\
		p_t-p_r & = \frac{r^2\left(2BM_1\frac{r^4}{r^5_b}-5M_1\frac{r^2}{r^3_b}-3\frac{M_2}{r_b}+3B\right)
			\left[\left(2B\frac{r^2}{r^2_b}-1\right)
			\left(M_1\frac{r^2}{r^3_b}+\frac{M_2}{r_b}\right) -B\right]}{4r^4_b\left(1-B\frac{r^2}{r^2_b}\right)^2 \left(1-M_1\frac{r^4}{r^5_b}-M_2\frac{r^2}{r^3_b}\right)}\\
		& -\frac{BM_1r^4}{r^7_b\left(3\sqrt{1-B}\sqrt{1-B\frac{r^2}{r_b}} +B\frac{r^2}{r^2_b}-1\right)}\,.
	\end{aligned}
\end{equation}
which shows that $p_r\left(r_b\right)=0$, for $B\neq 1$.

The conditions associated with the viability of this spacetime as a theoretical model 
for the interior of a compact star include: (i) the regularity and positivity of the 
metric coefficients, (ii) the regularity and signs of the matter potentials in 
relation to the energy conditions, (iii) the correct sign and value of the radial and 
tangential speed of sound, and (iv) the junction conditions to match the solution to 
the external vacuum.

Unfortunately, given the complexity of the expressions involved, the constraints on the parameters for which conditions (i)--(iv) are verified are too long and complicated to report in full here, but are relatively straightforward to compute. We present only some of these conditions explicitly and support our conclusions regarding the viability of this solution with plots for specific values of the parameters. We will focus in particular on the absence of horizons and singularities in the metric coefficients and the thermodynamic potentials, and the positivity of the matter energy density, radial pressure, and total mass of the stellar object, namely conditions (i) and (ii).

The metric coefficients and the thermodynamic quantities are regular for $r\in[0,r_b]$, if $B<\frac{8}{9}$ and
\begin{equation}
\begin{split}
& M_2 \leq 0\quad M_1>-M_2\quad r_b> M_1+M_2\\
& M_2>0 \quad -M_2<M_1<-\frac{M_2}{2}\quad r_b\geq -\frac{M_2^2}{4M_1}\\
& M_2>0 \quad -\frac{M_2}{2}\leq M_1<0\quad r_b> M_1+M_2\\
& M_2>0 \quad M_1>0\quad r_b > M_1+M_2
\end{split}
\end{equation}
In addition, the energy density  will be positive between the origin, if 
\begin{equation}
\begin{split}
& M_2>0\quad M_1\geq -M_2
\end{split}
\end{equation} 
which also guarantees that $r_b>0$.
Combining the above results and further requiring that the radial pressure is positive, we have the conditions (i) and (ii) are verified for all $r\in[0,r_b]$, if $B<\frac{8}{9}$, $M_2>0$ and
\begin{equation}
\begin{split}
&-\frac{3 M_2}{5}<M_1\leq -\frac{M_2}{2}\land r_b\geq-\frac{M_2^2}{4 M_1}\\
&-\frac{M_2}{2}<M_1<0\land r_b > M_1+M_2\,.
\end{split}
\end{equation}

\begin{figure}[h]
	\centering
	\includegraphics[width=1\columnwidth]{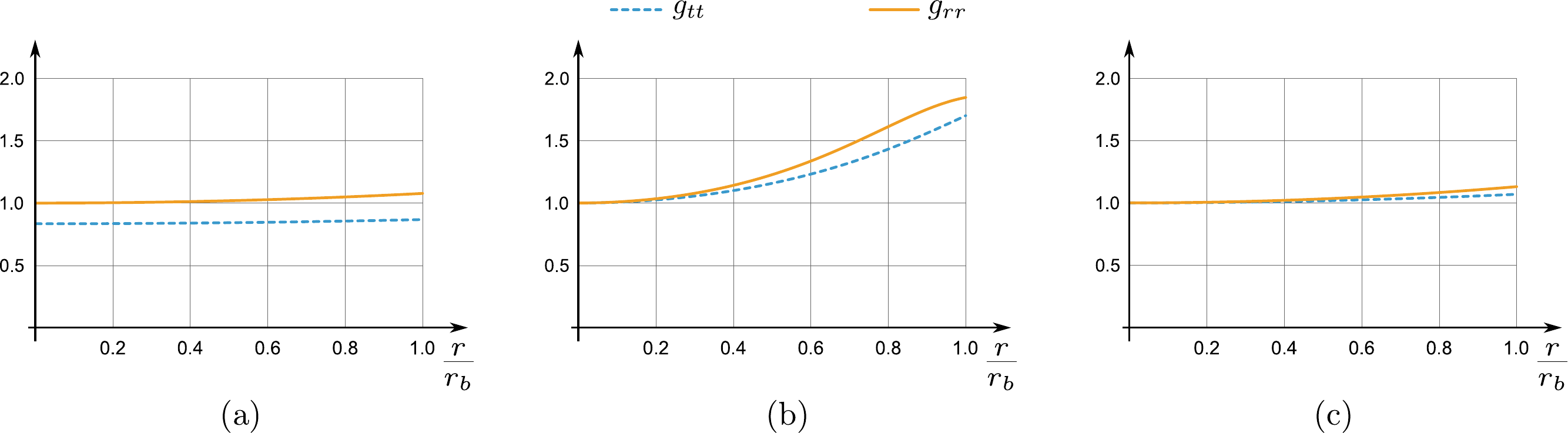}
	\caption{
		\label{Figure:Non_constant_solution_metric_plots}
		Non-trivial metric components for the 
		solution~\eqref{Sol_Var_mu_Par}, for various
		values of the parameters.
		In all cases, we have set the constant $A$ such that $g_{tt}(0)=1$.
		In (a) $M_1=-\frac{1}{480}$, $M_2=\frac{1}{20}$, $B=\frac{1}{12}$, and $r_b=\frac{2}{3}$.
		(b)  $M_1=-\frac{9}{20}$, $M_2=1$, $B=-1$, and $r_b=\frac{6}{5}$.
		(c)  $M_1=-\frac{32}{3645}$, $M_2=\frac{104}{1215}$, $B=-\frac{32}{3645}$, and $r_b=\frac{2}{3}$.	
	}
\end{figure}

\begin{figure}[h]
	\centering
	\includegraphics[width=1\columnwidth]{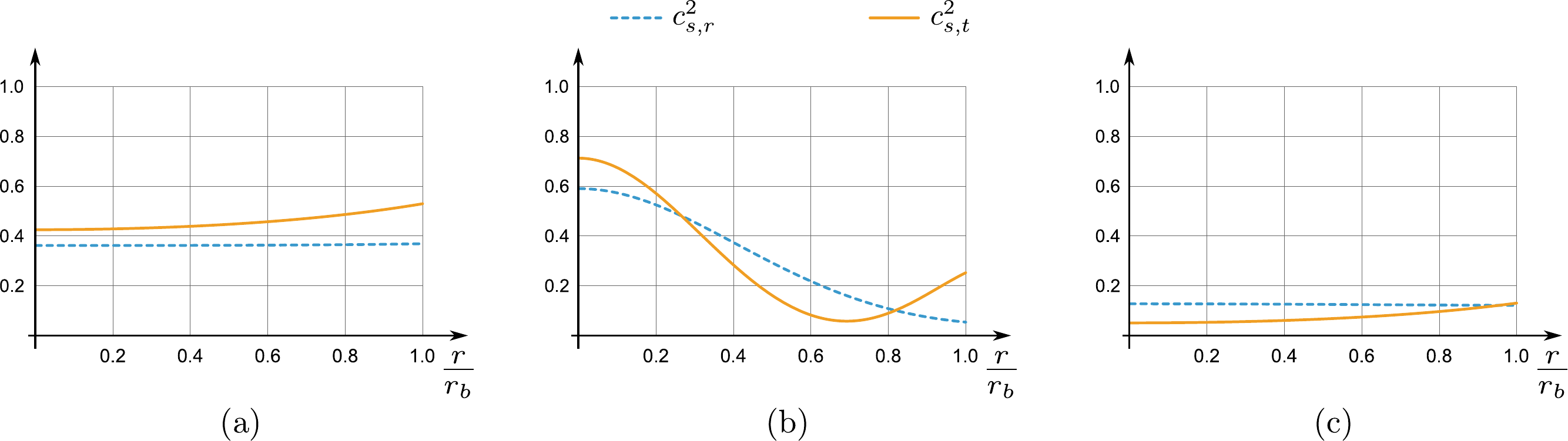}
	\caption{
		\label{Figure:Non_constant_solution_sound_plots}
		Radial profile of the radial, $c^2_{s,r}$, and tangential, $c^2_{s,t}$, sound speeds for the 
		solution~\eqref{Sol_Var_mu_Par}, for various
		values of the parameters.
		In (a) $M_1=-\frac{1}{480}$, $M_2=\frac{1}{20}$, $B=\frac{1}{12}$, and $r_b=\frac{2}{3}$.
		(b)  $M_1=-\frac{9}{20}$, $M_2=1$, $B=-1$, and $r_b=\frac{6}{5}$.
		(c)  $M_1=-\frac{32}{3645}$, $M_2=\frac{104}{1215}$, $B=-\frac{32}{3645}$, and $r_b=\frac{2}{3}$.	
	}
\end{figure}

An example of the features of solution~\eqref{Sol_Var_mu_Par} is given in Figures \ref{Figure:Non_constant_solution_metric_plots}--\ref{Figure:Non_constant_solution_sound_plots}. Interestingly, some combinations of parameters that satisfy the criteria listed above yield a remarkably close-to-isotropic pressure profile. Similar quasi-isotropic solutions were already found in Ref. \cite{Carloni:2017rpu}.

\begin{figure}[h]
	\centering
	\includegraphics[width=1\columnwidth]{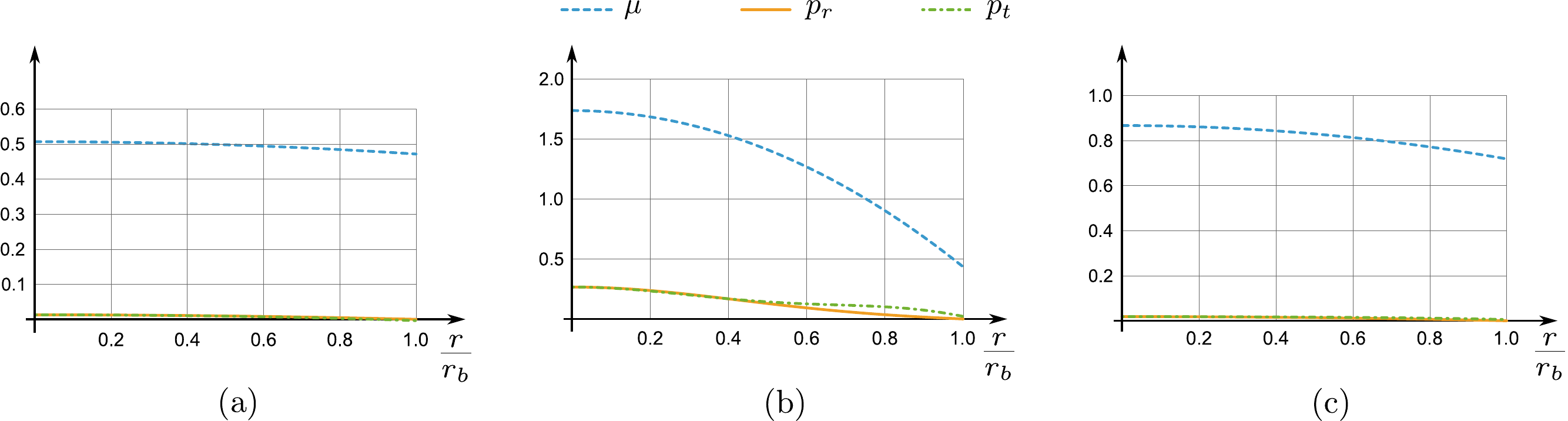}
	\caption{
		Radial profile of the energy density, radial and tangential pressures for the 
		solution~\eqref{Sol_Var_mu_Par}, for various
		values of the parameters.
		In (a) $M_1=-\frac{1}{480}$, $M_2=\frac{1}{20}$, $B=\frac{1}{12}$, and $r_b=\frac{2}{3}$.
		(b)  $M_1=-\frac{9}{20}$, $M_2=1$, $B=-1$, and $r_b=\frac{6}{5}$.
		(c)  $M_1=-\frac{32}{3645}$, $M_2=\frac{104}{1215}$, $B=-\frac{32}{3645}$, and $r_b=\frac{2}{3}$.	
	}
\end{figure}

\section{Conclusions }\label{Concl}
Using the 1+1+2 formalism to write a covariant version of the TOV equations, we have constructed and employed a new class of generating theorems to explore the features of solutions of the equations, and derive new anisotropic solutions of the Einstein field equations.

The existence of generating theorems can be understood by looking at the structure of the
covariant TOV equations. Indeed, the TOV equations are obtained by reducing the 1+1+2 
equations to two ordinary differential equations, along with a constraint. The 
existence of such a constraint is the key point underlying the generating systems. It 
shows that, given a solution,  another combination of the variables that satisfies 
the constraint can also be a solution. Clearly, there is no guarantee that the new 
solution can be successfully connected to any realistic physical system. Still, from a 
mathematical point of view, it reveals an interesting structure of the solution space 
for the TOV equations.

One might wonder what the general implications of the existence of generating theorems are in 
terms of the nature of the solutions of the TOV equations.  Since the generating 
theorems change the form of the thermodynamic variables, and therefore the equation 
of state, in general, the solutions connected by these theorems link different 
physical systems. The case of the solutions of the TOV equations with constant 
density is particularly interesting in this respect, as it is connected to the 
uniqueness of the incompressible limit. In the case of the Interior 
Schwarzschild solution, applying generating theorems that preserve isotropy and the constant 
density profile simply led to reparametrization of the solution. In this sense, the Interior Schwarzschild solution that represents the limiting case for an isotropic stellar object can be considered 
unique.   

The situation is different for non-isotropic compact objects, as the additional degrees of freedom allow non-trivial deformations even when imposing a constant density.
Indeed, applying a new generating theorem that combines isotropic and anisotropic pressures, starting from the Interior Schwarzschild solution, we derived infinitely many formal solutions, either by imposing a constant density or by considering deformations of the energy density that can be expressed as polynomials or power series.

In this article, we have focused on two closed-form solutions: the first, sourced by an anisotropic fluid with constant density, and the second, in which the energy density grows as quadratically with the circumferential radius.

Focusing our analysis on the new constant-density solution, we have found that this solution can represent the interior of a compact stellar object by evaluating the parameter ranges for which the geometry is non-singular and the source fluid satisfies the weak and strong energy conditions, for any value of the compactness parameter, up to the black hole limit.

Moreover, we have shown that through the new solution, the Interior Schwarzschild spacetime can be continuously connected with the Florides spacetime. The new solution also has, as a particular case, a non-trivial spacetime sourced by fluids with identically zero energy density but non-zero pressure. This is, to our knowledge, the first example of a solution that models ghost stars without curvature singularities. We have also discussed the possibility of matching these solutions to either an exterior asymptotically flat branch of the vacuum Schwarzschild solution or to Minkowski spacetime, either via a thin matter shell or smoothly.

The results presented here show that there exist at least two different families of constant density anisotropic solutions: the Bowers-Liang one and the Solution~\eqref{Constant_density_sol_eq:metric_rewritten} we have described above.
The existence of different branches of solutions for anisotropic fluids with constant energy density poses the question of which one should be used to model the incompressible limit for anisotropic compact stellar objects.
We have shown that both solutions can be extended to the black hole limit without violating the weak and strong energy conditions. However, for the Bowers-Liang solution, this requires that the radial pressure remain negative inside the star and that the tangent pressure be negative in a region around the center.
Nonetheless, the key test for discriminating between the two options is the analysis of the stability of the solutions under perturbations.  This topic will be investigated in more detail in future work.

The non-constant density solution turned out to also have some interesting features 
for the description of relativistic stars. We found, albeit in a somewhat heuristic 
manner, that these solutions can be characterized as quasi-isotropic, in the sense 
that the radial and tangential pressures differ only slightly. Otherwise, they can be 
shown to satisfy all the basic conditions to represent physical systems. Naturally, 
additional key aspects to be evaluated are the stability of such solutions  and the 
compatibility of the equation of state of the solution with those typical of nuclear 
matter.  None of these tasks is elementary, and they will be addressed elsewhere. 
Another remark concerns the range of applicability of these solutions. Even in cases 
where the thermodynamic quantities violate the requirements for standard fluids, the 
solutions we have found can still be useful for relativistic stars in modified gravity. It 
is well known that these theories can often be recast as general relativity plus some 
effective fluid. These fluids need not satisfy the standard energy conditions and, as such, 
can be compatible with our solution over a wider range of parameters.

\begin{acknowledgments}
	PL thanks the Funda\c{c}\~{a}o para a Ci\^{e}ncia e Tecnologia (FCT), Portugal, for the financial support to the Center for Astrophysics and Gravitation through grant No. UID/PRR/00099/2025 and grant No. UID/00099/2025.
\end{acknowledgments}


\begin{thebibliography}{99}

\bibitem{Herrera:1997plx}
L.~Herrera, and N.~O.~Santos,
``Local anisotropy in self-gravitating systems'',
Phys. Rept. \textbf{286}, 53 (1997).

\bibitem{Herrera_Barreto_2013}
L. Herrera, and W. Barreto, ``General
relativistic polytropes for anisotropic matter: The general formalism
and applications'', Phys. Rev. D \textbf{88}, 084022 (2013).

\bibitem{Herrera:2015vca}
L.~Herrera, E.~Fuenmayor, and P.~Le{\'o}n,
``Cracking of general relativistic anisotropic polytropes'',
Phys. Rev. D \textbf{93}, 024047 (2016).

\bibitem{Herrera:2007kz}
L.~Herrera, J.~Ospino, and A.~Di Prisco,
``All static spherically symmetric anisotropic solutions of Einstein's equations'',
Phys. Rev. D \textbf{77}, 027502 (2008).

\bibitem{Chaisi:2006sc}
M.~Chaisi, and S.~D.~Maharaj,
``A New algorithm for anisotropic solutions'',
Pramana \textbf{66}, 313  (2006).

\bibitem{Harko:2002pxr}
T.~Harko, and M.~K.~Mak,
``An Exact Anisotropic Quark Star Model'',
Chin. J. Astron. Astrophys. \textbf{2}, 248 (2002).

\bibitem{Harko:2003pxr}
M.~K.~Mak, and T.~Harko,
``Anisotropic stars in general relativity'',
Proc. Roy. Soc. Lond. A \textbf{459}, 393 (2003).

\bibitem{Hoffberg:1970vqj}
M.~Hoffberg, A.~E.~Glassgold, R.~W.~Richardson, and M.~Ruderman,
``Anisotropic Superfluidity in Neutron Star Matter'',
Phys. Rev. Lett. \textbf{24}, 775 (1970).

\bibitem{Rago:1991qe}
H.~Rago, ``Anisotropic spheres in general relativity'', IC-91-13.

\bibitem{Richardson:1972xn}
R.~W.~Richardson,
``Ginzburg-landau theory of anisotropic superfluid neutron-star matter'',
Phys. Rev. D \textbf{5}, 1883 (1972).

\bibitem{Viaggiu:2008yf}
S.~Viaggiu,
``Modelling usual and unusual anisotropic spheres'',
Int. J. Mod. Phys. D \textbf{18}, 275 (2009).

\bibitem{LIGOScientific:2016aoc}
B.~P.~Abbott \textit{et al.} [LIGO Scientific and Virgo],
``Observation of Gravitational Waves from a Binary Black Hole Merger'',
Phys. Rev. Lett. \textbf{116}, 061102 (2016).

\bibitem{Cadogan:2024mcl}
T.~Cadogan, and E.~Poisson,
``Self-gravitating anisotropic fluids. I: Context and overview'',
Gen. Rel. Grav. \textbf{56}, 118 (2024).

\bibitem{Cadogan:2024ohj}
T.~Cadogan, and E.~Poisson,
``Self-gravitating anisotropic fluid. II: Newtonian theory'',
Gen. Rel. Grav. \textbf{56}, 119 (2024).

\bibitem{Cadogan:2024ywc}
T.~Cadogan, and E.~Poisson,
``Self-gravitating anisotropic fluid. III: Relativistic theory'',
Gen. Rel. Grav. \textbf{56}, 120 (2024).

\bibitem{Abellan:2020nkl}
G.~Abellan, E.~Fuenmayor, E.~Contreras, and L.~Herrera,
``The general relativistic double polytrope for anisotropic matter'',
Phys. Dark Univ. \textbf{30}, 100632 (2020).

\bibitem{Thirukkanesh:2018hfy}
S.~Thirukkanesh, F.~C.~Ragel, R.~Sharma, and S.~Das,
``Anisotropic generalization of well-known solutions describing relativistic self-gravitating fluid systems: an algorithm'',
Eur. Phys. J. C \textbf{78}, 31 (2018).

\bibitem{Estrada:2018zbh}
M.~Estrada, and F.~Tello-Ortiz,
``A new family of analytical anisotropic solutions by gravitational decoupling'',
Eur. Phys. J. Plus \textbf{133}, 453 (2018).

\bibitem{Singh:2016mqs}
K.~N.~Singh, N.~Pant, and M.~Govender,
``Some analytic models of relativistic compact stars'',
Indian J. Phys. \textbf{90}, 1215 (2016).

\bibitem{Maurya:2016oml}
S.~K.~Maurya, Y.~K.~Gupta, S.~Ray, and D.~Deb,
``Generalised model for anisotropic compact stars'',
Eur. Phys. J. C \textbf{76}, 693 (2016).

\bibitem{Maurya:2017gth}
S.~K.~Maurya,
``Relativistic modeling of compact stars for anisotropic matter distribution'',
Eur. Phys. J. A \textbf{53}, 89 (2017).

\bibitem{Ivanov:2017kyr}
B.~V.~Ivanov,
``Analytical study of anisotropic compact star models'',
Eur. Phys. J. C \textbf{77}, 738 (2017).

\bibitem{Jasim:2016cmk}
M.~K.~Jasim, S.~K.~Maurya, Y.~K.~Gupta, and B.~Dayanandan,
``Well behaved anisotropic compact star models in general relativity'',
Astrophys. Space Sci. \textbf{361}, 352 (2016).

\bibitem{Maurya:2016ecj}
S.~K.~Maurya, Y.~K.~Gupta, B.~Dayanandan, and S.~Ray,
``A new model for spherically symmetric anisotropic compact star'',
Eur. Phys. J. C \textbf{76}, 266 (2016).

\bibitem{Bhar:2017ynp}
P.~Bhar, K.~N.~Singh, and T.~Manna,
``A new class of relativistic model of compact stars of embedding class I'',
Int. J. Mod. Phys. D \textbf{26}, 1750090 (2017).

\bibitem{Ovalle:2017fgl}
J.~Ovalle,
``Decoupling gravitational sources in general relativity: from perfect to anisotropic fluids'',
Phys. Rev. D \textbf{95}, 104019 (2017).

\bibitem{Gabbanelli:2018bhs}
L.~Gabbanelli, {\'A}.~Rinc{\'o}n, and C.~Rubio,
``Gravitational decoupled anisotropies in compact stars'',
Eur. Phys. J. C \textbf{78}, 370 (2018).

\bibitem{Ovalle:2017wqi}
J.~Ovalle, R.~Casadio, R.~da Rocha, and A.~Sotomayor,
``Anisotropic solutions by gravitational decoupling'',
Eur. Phys. J. C \textbf{78}, 122 (2018).

\bibitem{Maurya:2015maa}
S.~K.~Maurya, Y.~K.~Gupta, S.~Ray, and B.~Dayanandan,
``Anisotropic models for compact stars'',
Eur. Phys. J. C \textbf{75}, 225 (2015).

\bibitem{Clarkson_Barrett_2003}
C. A. Clarkson, and R. K. Barrett,
``Covariant perturbations of Schwarzschild black holes'',
Class. Quant. Grav. \textbf{20} ,3855 (2003).

\bibitem{Clarkson_2007}
C. A. Clarkson, ``A covariant
approach for perturbations of rotationally symmetric spacetimes'',
Phys. Rev. D \textbf{76}, 104034 (2007).

\bibitem{Betschart_Clarkson_2004}
G. Betschart, and C. A. Clarkson,
``Scalar field and electromagnetic perturbations on
Locally Rotationally Symmetric spacetimes'', Class.
Quant. Grav. \textbf{21}, 5587 (2004).

\bibitem{Luz_Carloni_2024a}
P. Luz, and S. Carloni, ``Gauge invariant
perturbations of static spatially compact LRS II spacetimes'', Class.
Quant. Grav. \textbf{41}, 235012 (2024).

\bibitem{Luz_Carloni_2024b}
P. Luz, and S. Carloni, ``Adiabatic radial
perturbations of relativistic stars: Analytic solutions to an old
problem'', Phys. Rev. D \textbf{110}, 084054 (2024).

\bibitem{Luz_Carloni_2024c}
P. Luz, and S. Carloni, ``Noncomoving
description of adiabatic radial perturbations of relativistic stars'',
Phys. Rev. D \textbf{110}, 084055 (2024).

\bibitem{Carloni:2017rpu}
S.~Carloni, and D.~Vernieri,
``Covariant Tolman-Oppenheimer-Volkoff equations. I. The isotropic case'',
Phys. Rev. D \textbf{97}, 124056 (2018).

\bibitem{Carloni:2017bck}
S.~Carloni, and D.~Vernieri,
``Covariant Tolman-Oppenheimer-Volkoff equations. II. The anisotropic case'',
Phys. Rev. D \textbf{97}, 124057 (2018).

\bibitem{Rahman}
S.~Rahman, and M.~Visser,
``Spacetime geometry of static fluid spheres'',
Class. Quant. Grav. \textbf {19}, 935 (2002).


\bibitem{Lake}
K.~Lake,
``All static spherically symmetric perfect fluid solutions of Einstein's
Equations'',
Phys. Rev. D \textbf{67}, 104015 (2003).


\bibitem{Martin}
D.~Martin, and M.~Visser,
``Algorithmic construction of static perfect fluid spheres'',
Phys. Rev. D textbf {69}, 104028 (2004).


\bibitem{Boonserm:2005ni} 
P.~Boonserm, M.~Visser, and S.~Weinfurtner,
``Generating perfect fluid spheres in general relativity'',
Phys. Rev. D \textbf{71}, 124037 (2005).
  
\bibitem{Boonserm:2006up} 
P.~Boonserm, M.~Visser, and S.~Weinfurtner,
``Solution generating theorems for the TOV equation'',
Phys. Rev. D \textbf{76}, 044024 (2007).
  
\bibitem{ExactSolutions}
H.~Stephani, D.~Kramer, M.~MacCallum, C.~Hoenselaers, and E.~Herlt , 
\emph{Exact solutions of Einstein's field equations} Cambridge University Press (2009).

\bibitem{Florides1974abc} 
P.~S.~Florides, "A new interior Schwarzschild solution",
Proc. Roy. Soc. Lon. A \textbf{337}, 1611 (1974).

\bibitem{Bowers_Liang_1974}
R. L. Bowers, and E. P. T. Liang, ``Anisotropic
Spheres in General Relativity'', Astrophys. J. \textbf{188}, 657
(1974).

\bibitem{Yagi_Yunes_2015}
K. Yagi, and N. Yunes, ``I-Love-Q anisotropically:
Universal relations for compact stars with scalar pressure anisotropy'',
Phys. Rev. D \textbf{91}, 123008 (2015).

\bibitem{Silva_et_al_2015}
H. O. Silva, C. F. B. Macedo, E. Berti,
and L. C. B. Crispino, ``Slowly rotating anisotropic neutron stars
in general relativity and scalar--tensor theory'', Class. Quantum
Grav. \textbf{32}, 145008 (2015).

\bibitem{Singh_et_aj_2019}
K. N. Singh, F. Rahaman, and A. Banerjee,
``Einstein\textquoteright s cluster mimicking compact star in the
teleparallel equivalent of general relativity'', Phys. Rev. D \textbf{100},
084023 (2019).

\bibitem{Olmo_et_al_2020}
G. J. Olmo, D. Rubiera-Garcia, and A. Wojnar,
``Stellar structure models in modified theories of gravity: Lessons
and challenges'', Phys. Rept. \textbf{876}, 1 (2020).

\bibitem{Becerra_2026}
L. M. Becerra, E. A. Becerra-Vergara, F. D.
Lora-Clavijo, and J. F. Rodriguez, ``Stability of anisotropic neutron
stars'', Phys. Rev. D \textbf{113}, 023009 (2026).


\bibitem{Herrera_Prisco_Ospino_2024}
L. Herrera, A. Di Prisco, and
J. Ospino, ``Ghost stars in general relativity'',
Symmetry \textbf{16},
562 (2024).

\bibitem{Zeldovich_Novikov_Book_1971}
Ya. B. Zeldovich, and I. D.
Novikov, \emph{Relativistic Astrophysics. Vol I. Stars and Relativity}
University of Chicago Press, Chicago, 1971.
\end{thebibliography}
\end{document}